\documentclass{article} 
\usepackage{style,times}

\usepackage{amsmath,amsfonts,bm}

\def\eqref#1{equation~\ref{#1}}

\def\1{\bm{1}}

\DeclareMathAlphabet{\mathsfit}{\encodingdefault}{\sfdefault}{m}{sl}
\SetMathAlphabet{\mathsfit}{bold}{\encodingdefault}{\sfdefault}{bx}{n}

\newcommand{\E}{\mathbb{E}}

\usepackage{hyperref}
\usepackage{url}
\usepackage{graphicx}
\usepackage{svg}
\usepackage{wrapfig}

\title{On Recovering Higher-order Interactions from Protein Language Models}
\author{Darin Tsui \& Amirali Aghazadeh \\
School of Electrical and Computer Engineering\\
Georgia Institute of Technology\\
Atlanta, GA 30332, USA \\
\texttt{\{darint,amiralia\}@gatech.edu} \\
}

\finalcopy 

\begin{document}

\maketitle

\begin{abstract}
Protein language models leverage evolutionary information to perform state-of-the-art 3D structure and zero-shot variant prediction. Yet, extracting and explaining \emph{all} the mutational interactions that govern model predictions remains difficult as it requires querying the entire amino acid space for $n$ sites using $20^n$ sequences, which is computationally expensive even for moderate values of $n$ (e.g., $n\sim10$). Although approaches to lower the sample complexity exist, they often limit the interpretability of the model to just single and pairwise interactions. Recently, computationally scalable algorithms relying on the assumption of sparsity in the Fourier domain have emerged to learn interactions from experimental data. However, extracting interactions from language models poses unique challenges: it's unclear if sparsity is always present or if it is the only metric needed to assess the utility of Fourier algorithms. Herein, we develop a framework to do a systematic Fourier analysis of the protein language model ESM2 applied on three proteins—green fluorescent protein (GFP), tumor protein P53 (TP53), and G domain B1 (GB1)—across various sites for 228 experiments. We demonstrate that ESM2 is dominated by three regions in the sparsity-ruggedness plane, two of which are better suited for sparse Fourier transforms. Validations on two sample proteins demonstrate recovery of all interactions with $R^2=0.72$ in the more sparse region and $R^2=0.66$ in the more dense region, using only 7 million out of $20^{10}\sim10^{13}$ ESM2 samples, reducing the computational time by a staggering factor of 15,000. All codes and data are available on our GitHub repository  \href{https://github.com/amirgroup-codes/InteractionRecovery}{https://github.com/amirgroup-codes/InteractionRecovery}.
\end{abstract}

\section{Introduction}

Recent advances in transformer-based deep learning models have leveraged evolutionary information to learn biological patterns in protein sequences. These models, encompassing up to 15 billion learnable parameters, are trained on amino acid sequences stored in databases such as UniProt ~\citep{Lin23, UniProt15}. In particular, masked language models have been demonstrated to achieve state-of-the-art performance in zero-shot variant effect and protein structure prediction without the need for explicit training~\citep{Meier2021, Brandes23}. Hence, it's widely believed that protein language models encapsulate representations that reflect the fundamental rules of biology and physics~\citep{Rives21, Rao20}. However, further applications of protein language models, e.g., for knowledge discovery, are hindered due to the challenge of interpreting the biological interactions that underlie their predictions.

In principle, if we wanted to learn the structural impact of variants underlying $n$ mutational sites in a protein, referred to as the region's landscape, we could query these language models on all possible $20^n$ mutational combinations (for all 20 standard amino acids). However, computational challenges would make such an endeavor nearly unrunnable at a large scale.  For instance, on four NVIDIA RTX A6000s, each sample takes about 0.01 seconds to compute. It would take $20^n \times 0.01 = 32000$ seconds, or around nine hours, to compute all possible combinations for $n=5$. However, even just increasing the length to $n=8$ would make the entire space take 194 years to complete. 

\begin{figure}[!t]
\begin{center}
\includegraphics[width=\textwidth]{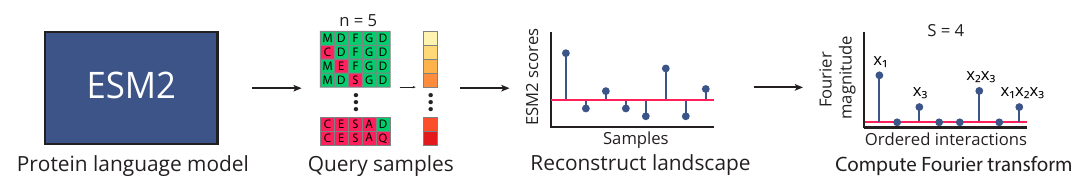}
\end{center}
\caption{Schematic of our ESM2 Fourier analysis framework. ESM2, a protein language model trained for masked language modeling, predicts amino acids at various positions. Using a fixed mutation, we query the entire combinatorial space for selected positions and compute the Fourier transform to recover important interactions.}
\label{fig:pipeline}
\end{figure}

Querying all possible combinations to learn a protein language model's landscape is beyond the capabilities of today's computers. Hence, we need a smarter and more efficient way to parse through and interpret the model's outputs with a lower sample complexity. Methods such as DeepSHAP and DeepLIFT assume interactions are locally additive in order to quantify mutational effects~\citep{Shrikumar17, Lundberg17}. However, biological interactions often exhibit local and global nonlinearities, or epistasis, due to physical interactions between amino acids and other biophysical properties~\citep{Starr16, Otwinowski18}. Other works, when given experimental data, attempt to rectify this by fitting a genotype-phenotype map. This mapping attempts to explicitly model mutational interactions in the data according to the user, such as accounting for all single and/or pairwise interactions~\citep{Otwinowski18, Tareen22}. Other frameworks, such as SQUID~\citep{Seitz23}, extend this approach by interpreting genomic deep neural networks in user-specified regions instead of experimental data. They fit the neural network output to biologically interpretable models to capture local interactions while maintaining nonlinearity~\citep{Seitz23}. However, both these methods restrict the interpretability of the model to up to pairwise interactions. As such, if the protein language model was actually relying on higher-order interactions to make its predictions, we would be losing the complexity of the original model in favor of preconceived assumptions.

One alternative approach that maintains the original model's complexity without compromising on interpretability involves converting the neural network's outputs to its spectral representation (Fourier transform). Here, we treat the neural network as a black box and do not assume anything about what the model learned. By transforming scores from a neural network to its Fourier representation, we encode such complex functions in terms of its mutational interactions~\citep{aghazadeh2020crisprl,Brookes22}. Hence, each Fourier coefficient represents the contribution of specific mutations on the prediction of the neural network~\citep{Aghazadeh21}. In the Boolean case, the Fourier transform is equivalent to the Walsh-Hadamard Transform (WHT), which we use interchangeably in this paper. 

While typically, Fourier-based algorithms require an exponential sample complexity to compute; sparse Fourier-based transforms have recently been developed to lower this barrier. By assuming that the Fourier transform is $S$-sparse, the sample complexity needed to recover the landscape scales \emph{linearly} instead of exponentially with $n$ (i.e., $O(Sn)$)~\citep{Erginbas23}. Yet, it is unclear whether sparsity consistently exists in the protein language Fourier domain or to what degree of sparsity is necessary for effective interaction recovery. Furthermore, it's also unclear if sparsity is the only metric needed to assess the utility of sparse Fourier transforms. 

Herein, we develop a framework to extract meaningful biological interactions from protein language model landscapes. Using ESM2~\citep{Lin23}, a protein masked language model, we perform a systematic WHT analysis on three proteins across various regions based on experimental interest and 3D structure (Figure \ref{fig:pipeline}). We characterize each Fourier transform by considering sparsity and introducing a novel metric, ruggedness, to capture the transform's higher-order interactions. We then perform a sparse Fourier transform on both sparse and rugged sites. Our findings suggest that the ESM2 landscape is highly dependent on the protein context sequence and the mutational sites. Furthermore, ESM2 is surprisingly dominated by regions of higher-order interactions, which give rise to three regions in the sparsity-ruggedness plane, two of which are well-suited for sparse Fourier algorithms. We conclude by extracting all interactions for two sample proteins in these regions.

\section{Methods}
    
Our main goal is to measure the full amino acid landscape of the ESM2 scores. However, given the computational cost of measuring $20^{n}$ combinations across all landscapes, we instead examine the scenario where each site is restricted to mutate to a specific amino acid—a total of $2^{n}$ combinations. We look to test whether taking several $2^{n}$ combinations over random amino acid mutations is a good proxy for $20^{n}$. We then look to conduct the full $20^{n}$ sparse Fourier transform on a sparse region, which would be impossible without using $2^{n}$ as a proxy.

\textbf{Notation.} We are interested in taking the WHT of the ESM2 protein landscape, that is, the vector $\textbf{x} \in \mathbb{R}^N$, containing $N=2^n$ scores corresponding to the output of ESM2 once being input with combinatorial ways one can mutate a protein sequence over $n$ mutational sites. Let $\mathbb{F}_2^n=\{0,1\}^n$ represent the $n$-dimensional column vector living in the binary representation $\{0,1\}$. Furthermore, let the vector $\textbf{m} \in \mathbb{F}_2^n$ represent the binary representation of $m \in [N]$, or in other words, $\textbf{m} = [m[1],...,m[n]]^T \in \mathbb{F}_2^n$. For each landscape, we have $2^n$ WHT coefficients, where each coefficient can be denoted as $F[\textbf{m}]$. Mathematically, we aim to learn a function $f(x)$ that maps mutations of the sequence to ESM scores. Using the WHT, we can define this function as a polynomial made up of interactions. For instance, if we observe the functional landscape $f(x_1, x_2, x_3) = 3x_1 + x_3 + 2x_2x_3 + x_1x_2x_3$, then we can conclude the landscape consists of two first-order, one second-order, and one third-order interaction, with the WHT coefficients 3, 1, 2, and 1. In this, we can create an interpretable metric of how mutations influence the ESM score. 

\textbf{Sparsity and ruggedness.} To quantify each ESM2 landscape, we measure sparsity in the WHT domain as well as a novel metric called ruggedness. Sparsity is the number of non-zero coefficients in the landscape's WHT. We find sparsity by computing the number of WHT coefficients larger than an empirical threshold to filter out noisy coefficients. Mathematically, we define sparsity up to $5^{th}$ order interactions, $S_5$, as
\begin{equation}
    S_5  = \frac{ \sum_{{\bf m\in\mathbb{F}_2^n}:sum({\bf m})\leq 5 \text{  and  } |F[\textbf{m}]| > \frac{\sigma^2}{c} }1}{ \sum_{{\bf m\in\mathbb{F}_2^n}:sum({\bf m})\leq 5} 1   } 
\end{equation}

where $\sigma^2$ is the variance of the WHT coefficients and $c$ is some constant. As interaction order increases, the chance an interaction represents a biologically meaningful mutation decreases. Hence, we purposely constrain our sparsity estimate to a maximum of $5^{th}$ orders. In addition, we define a new metric, ruggedness, as the expected value of the order of interactions with respect to the square of the WHT coefficients: 
\begin{equation}
    \text{Ruggedness} = \E_{k} [ k ] \approx \sum_{{\bf m\in\mathbb{F}_2^n}:sum({\bf m})\leq 5} kF^2[\textbf{m}],\text{ } k = sum({\bf m})
\end{equation}
Ruggedness can be interpreted as the weighted average of the order ($k$) among the interactions: if ruggedness is close to $k$, the landscape is dominated by $k^{th}$ order interactions. To run sparse Fourier transforms, we desire sparse and rugged landscapes. 

\section{Results}

\subsection{Experimental Design}

\textbf{Protein sequence data.} We examine the ESM2 landscapes of three proteins: green fluorescent protein (GFP), tumor protein P53 (TP53), and G domain B1 (GB1)~\citep{Poelwijk19, Giacomelli18, Olson14}. These proteins encompass wide ranges of sparsity and ruggedness. We analyze the landscapes of five regions along each protein based on their 3D structure and experimental importance. Amino acid sites are randomly selected to form two secondary structures and two random coil regions. Additionally, one region was also comprised of sites previously investigated in experimental studies. All regions are comprised of $n=10$ sites each, with the exception of $13$ sites experimentally studied in GFP~\citep{Poelwijk19}. To compute the ESM scores, we utilize an implementation from~\citep{Brandes23}. We also compute the ESM scores given an implementation from~\citep{Meier2021}, which we leave in the Appendix.

\begin{figure}[!t]
\vspace{-1.5cm}
\begin{center}
\includegraphics[width=0.9\textwidth]{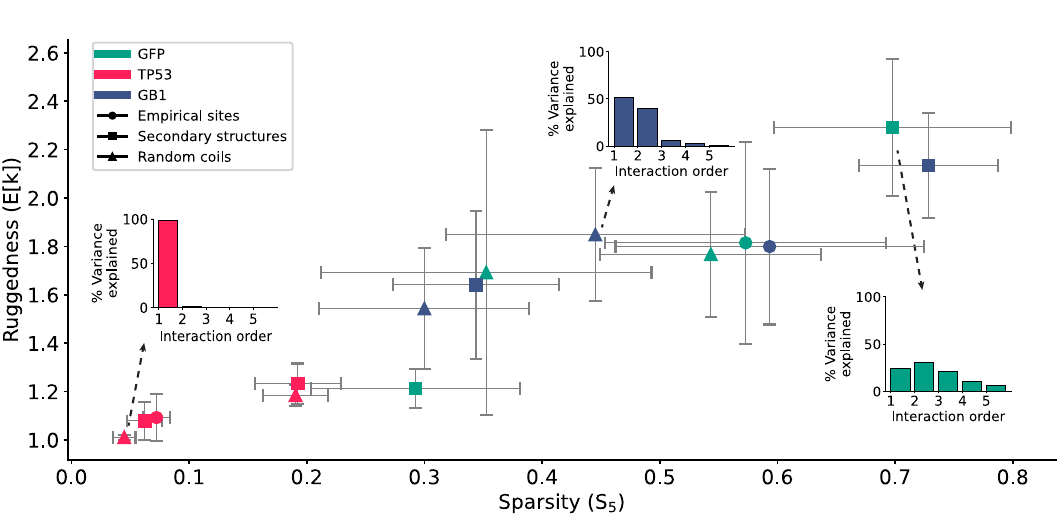}
\end{center}
\vspace{-0.2cm}
\caption{Ruggedness and sparsity across different ESM2 landscapes for GFP, TP53, and GB1. Site selection was based on experimental literature or random sampling from secondary structures and random coils. Our results demonstrate the presence of higher-order interactions, context sequence dependence of sparsity and ruggedness, and identification of regions for sparse Fourier transform.}
\vspace{-0.15cm}
\label{fig:landscapes}
\end{figure}

\textbf{Protein language model landscapes.} For each region, we query ESM2 $2^n$ times to calculate the WHT, sparsity, and ruggedness. Each site was only allowed to mutate to a randomly chosen fixed amino acid, which was randomly chosen. A total of 19 trials were run per region with different choices of mutated amino acids (Figure \ref{fig:landscapes}). 

\subsection{Assessment of Landscapes}

We assess the sparsity and ruggedness of 228 ESM2 landscapes across three proteins. Our systematic analysis of ESM2 reveals distinct characteristic patterns in the perceived interactions among amino acids. Herein, we summarized the key results.

\textbf{Presence of higher-order interactions.} Our analysis demonstrates the surprising presence of higher-order ($k\geq2$) interactions in ESM2 predictions for several proteins. While TP53 landscapes are largely explained by first-order interactions, landscapes from GFP and GB1 are dominated by pairwise and third-order interactions (with an average ruggedness of $1.8 \pm 0.4$). It would require another study to examine whether all or any of the extracted interactions are biological, albeit this observation seems to contrast with some studies that argue for only the interactions of up to second order to be physically meaningful~\citep{Figliuzzi18}.

\textbf{Context sequence and amino acid type dependence of sparsity and ruggedness.} Figure~\ref{fig:landscapes} demonstrates the strong dependence of sparsity and ruggedness in ESM2 landscapes with the context sequence around the mutations (i.e., the protein). TP53, on average, had the most sparse ($0.1 \pm 0.1$) and least rugged ($1.1 \pm 0.1$) Fourier landscapes. This is contrasted with GFP and GB1, which had denser ($0.5 \pm 0.2$ and $0.5 \pm 0.2$) and more rugged ($1.8 \pm 0.5$ and $1.8 \pm 0.3$) landscapes. Even within the same protein, ESM2 landscapes exhibit different characteristics. For instance, the sites sampled from GB1 exhibit Fourier transforms that are both sparse and not rugged while also having some of the most dense and rugged transforms. Sparsity and ruggedness can vary anywhere from $0.2$ to $0.8$ and $1.1$ to $2.5$, respectively. Figure~\ref{fig:landscapes} further demonstrates the strong dependence of sparsity and ruggedness of ESM2 landscapes with the type of amino acids at the same mutational sites. On average, the sparsity and ruggedness of landscapes vary $65.4\%$ and $30.4\%$, respectively, from their mean value. Our stratified Fourier analysis of landscapes across sites residing on secondary structures against random coils did not show correlations, perhaps due to the size of the experiments.

\begin{figure}[t]
\begin{center}
\vspace{-1cm}
\includegraphics[width=0.85\textwidth]{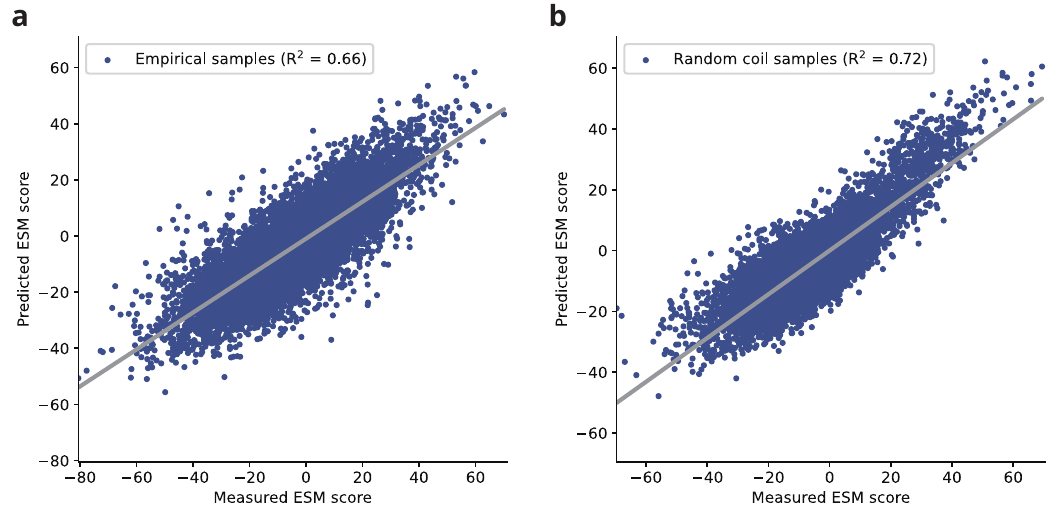}
\end{center}
\caption{Scatter plot of the predicted ESM scores using recovered Fourier coefficients on ten empirical (\textbf{a}) and random coil sites (\textbf{b}) from GB1 over the entire amino acid space. Sparse Fourier transforms recover most interactions with an NMSE of 0.32 ($R^2 = 0.66)$ and 0.26 ($R^2 = 0.72$), respectively, highlighting the recovery of interactions in ESM2 in sparse and rugged landscapes.}
\vspace{-0.2cm}
\label{fig:samples}
\end{figure}

\textbf{Recovery of higher-order interactions in the full amino acid space.} Sparsity and ruggedness can inform us about the strategy to recover higher-order interactions. In landscapes dominated by first-order interactions, one could reasonably query all possible single mutations and create a linear model to explain ESM2. On the other hand, in landscapes that are dense and rugged, querying all possible single mutations would not be sufficient. In this case, assumptions of sparsity are also violated, so it would be difficult to recover all interactions. However, when landscapes are sparse and rugged, we can recover higher-order interactions using sparse Fourier transforms. We accordingly identify three regions in the sparsity-rugged plane for ESM2. The first region, from sparsity values $0$ to $0.25$, represents protein landscapes where applying sparse Fourier transforms would be an easy problem—interactions can be recovered using a linear model. The second, from sparsity values $0.25$ to $0.5$, and the third regions, from sparsity values $0.5$ to $1$, signify areas where sparse Fourier transforms are applicable with ascending difficulties. We apply one such sparse Fourier algorithm, $q$-SFT, on the $20^{10}$ landscape using sites from GB1's random coils (region 2) and experimental sites from ~\citep{Olson14} (region 3)~\citep{Erginbas23}. To measure the accuracy of the recovered interactions, we take $10,000$ random samples from ESM2 and compute the normalized mean-squared error (NMSE), defined as $\text{NMSE} = \frac{\| \hat{f} - f \|^2}{\| f \|^2}$, where $\hat{f}$ is the reconstructed ESM score vector using the recovered Fourier coefficients, and $f$ is the ESM score vector. $q$-SFT recovers the topmost mutational interactions with a $0.32$ and $0.26$ NMSE, respectively (Figure \ref{fig:samples}). Of the interactions recovered in the empirical sites, $92.2\%$ and $7.8\%$ of the variance are explained by first and second-order interactions, respectively. In the random coil sites, $q$-SFT predominantly recovered first-order interactions: $99.2\%$ and $0.8\%$ of the variance are explained by first and second-order interactions, respectively. 

We note that using a naive approach of taking $20^{10}$ samples to recover all interactions from ESM2 would take $3247$ years on our server. $q$-SFT leverages sparsity in the Fourier domain and drastically reduces the running time by only requiring us to take $7,040,000$ samples from ESM2, less than $0.001\%$ of the entire space. At this scale, conventional sparse recovery algorithms such as LASSO that do not leverage the structures in Fourier transform will not even run on our server. 

\section{Discussion}

\textbf{First-order interaction dominance.} Our result in Figure \ref{fig:samples} demonstrates that higher-order interactions are recoverable using sparse Fourier transforms with an NMSE lower than $0.32$. In both test cases, the majority of interactions recovered were first-order, and $q$-SFT was not able to recover interactions higher than second-order. Yet, in the case of the full Fourier transform with combinatorial samples, we observe that a significant fraction of variance is explained by third-order interactions and above ($3\%$ for the random coils and $10\%$ for the empirical sites). While $q$-SFT  recovers the topmost interactions, when used in recovery problems with high noise and lower sparsity, it misses Fourier coefficients with higher order, which naturally appear with lower energy. Increasing the number of samples always improves recovery performance; however, the strategy is constrained by the available computing resources.

\textbf{Biological implications.} In this study, we focused on extracting predictive high-order interactions from protein language models. This remains an interesting question: what fraction of those high-order interactions are causal and biologically relevant? Answering this question would require a separate study~\citep{sapoval2022current}. Our framework, however, is the very step in addressing this fundamental question. From a broader perspective, we anticipate that our framework of opening the black box of protein language models will enable new scientific discoveries, from identifying new disease-causing mutations to engineering novel proteins.

\bibliography{paper}
\bibliographystyle{paper}

\newpage
\appendix
\section{Landscapes Across Different ESM Scores}

\begin{figure}[!h]
\begin{center}
\includegraphics[width=0.9\textwidth]{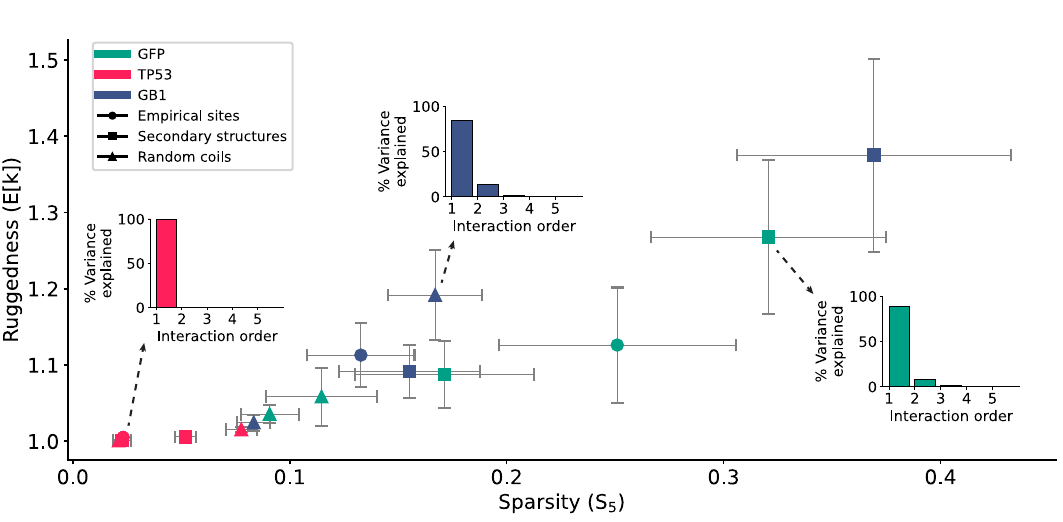}
\end{center}
\vspace{-0.2cm}
\caption{Ruggedness and sparsity across different ESM2 landscapes for GFP, TP53, and GB1 using ESM scores from~\citet{Meier2021}. While landscapes from~\citet{Meier2021} tend to be less rugged and more sparse than landscapes from~\citet{Brandes23}, both ESM scores demonstrate the presence of higher-order interactions.}
\vspace{-0.15cm}
\label{fig:landscapes_1}
\end{figure}

Figure \ref{fig:landscapes_1} illustrates the ruggedness-sparsity plot across different ESM2 landscapes using scores from~\cite{Meier2021}. Results present in the main text are calculated using scores from~\cite{Brandes23}. 

In both scenarios, we demonstrate the presence of higher-order interactions in GFP and GB1.~\citet{Meier2021} landscapes are less rugged ($1.1 \pm 0.1$ compared to $1.6 \pm 0.5$) and more sparse ($0.1 \pm 0.1$ compared to $0.4 \pm 0.2$) than~\citet{Brandes23}. On average, TP53 had the most sparse ($0.04 \pm 0.02$) and least rugged ($1.0 \pm 0.0$) Fourier landscapes. GFP and GB1 had denser ($0.2 \pm 0.1$ and $0.2 \pm 0.1$) and more rugged landscapes ($1.1 \pm 0.1$ and $1.2 \pm 0.1$), respectively. 

\end{document}